\documentclass[journal  =  jpclcd, manuscript=letter, layout=onecolumn]{achemso}
\usepackage{color}
\usepackage{amssymb}
\usepackage[english]{babel}
\usepackage{amsmath}
\usepackage{amsthm}
\usepackage{amsfonts}
\usepackage[version=3]{mhchem} 

\title{Microscopic Origin of the Hofmeister Effect in  Gelation Kinetics of Colloidal Silica}

    \author{Marte van der Linden*}
\affiliation{Cavendish Laboratory, University of Cambridge, U.~K.}

    \author{{Breannd\'{a}n} O. {Conch\'{u}ir}*}
\affiliation{Cavendish Laboratory, University of Cambridge, U.~K.}

\author{Elisabetta Spigone}
\affiliation{Cavendish Laboratory, University of Cambridge, U.~K.}

  \author{Arun Niranjan }
\affiliation{Cavendish Laboratory, University of Cambridge, U.~K.}

     \author{Alessio Zaccone}
\affiliation{Cavendish Laboratory, University of Cambridge, U.~K. and
Physics  Department  and  Institute  for  Advanced  Study,
Technische  Universit\"{a}t  M\"{u}nchen,    Garching,  Germany.}

  \author{Pietro Cicuta}
\affiliation{Cavendish Laboratory, University of Cambridge, U.~K.}
  \email{pc245@cam.ac.uk}

\begin{document}

\begin{tocentry}
\includegraphics[width=5.1cm]{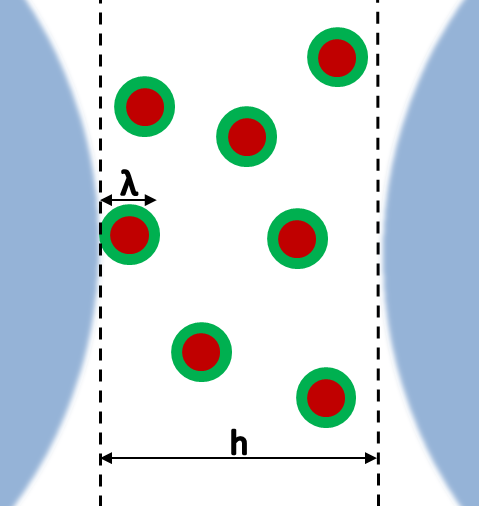}
\end{tocentry}

\begin{abstract}
{The gelation kinetics of silica nanoparticles  is a central process in physical chemistry, yet not fully understood.
 Gelation times are measured to increase by over four orders of magnitude,  simply changing the
 monovalent salt species from CsCl to LiCl. This striking effect has no microscopic explanation within current paradigms.
The trend is   consistent with the Hofmeister series,  pointing to short-ranged solvation effects not included in the standard colloidal (DLVO) interaction potential.  By implementing
a simple form for short-range repulsion  within a model that relates the gelation
time-scale to the colloidal interaction forces, we are able to explain  the many orders of
magnitude difference in the gelation times at fixed salt concentration. The  model allows  to estimate the magnitude of the non-DLVO hydration forces, which dominate the
 interparticle interactions at the length-scale of the hydrated ion diameter. This opens the possibility of finely tuning the gelation time-scale of nanoparticles by just adjusting the background electrolyte species.}
\end{abstract}




\vspace{2cm}

The total interaction potential in acqueous suspensions of charged colloidal particles is often  taken as the sum of the van der Waals attraction
and a simple approximation of the  electrostatic {double-layer} repulsion, forming
 the classical {Derjaguin-Landau-Verwey-Overbeek (DLVO)} potential~\cite{hunter86book}.
 The electrostatic repulsion
can  be overcome by adding salt,
thus increasing the screening of electrostatic repulsion, and lowering the energy barrier against
aggregation~\cite{Israelachvili2011a}. Colloid particles then typically aggregate into clusters, which grow over
time with the possibility of forming a sample-spanning network.
 Silica particles, and specifically Ludox,  have been a classical model
system for  colloidal physical chemistry, and  also have a key place in industrial processing, coatings, ink receptive papers, metal
casting, refractory products, and
catalysts.

\begin{figure*}[t!]
\includegraphics[width=14cm]{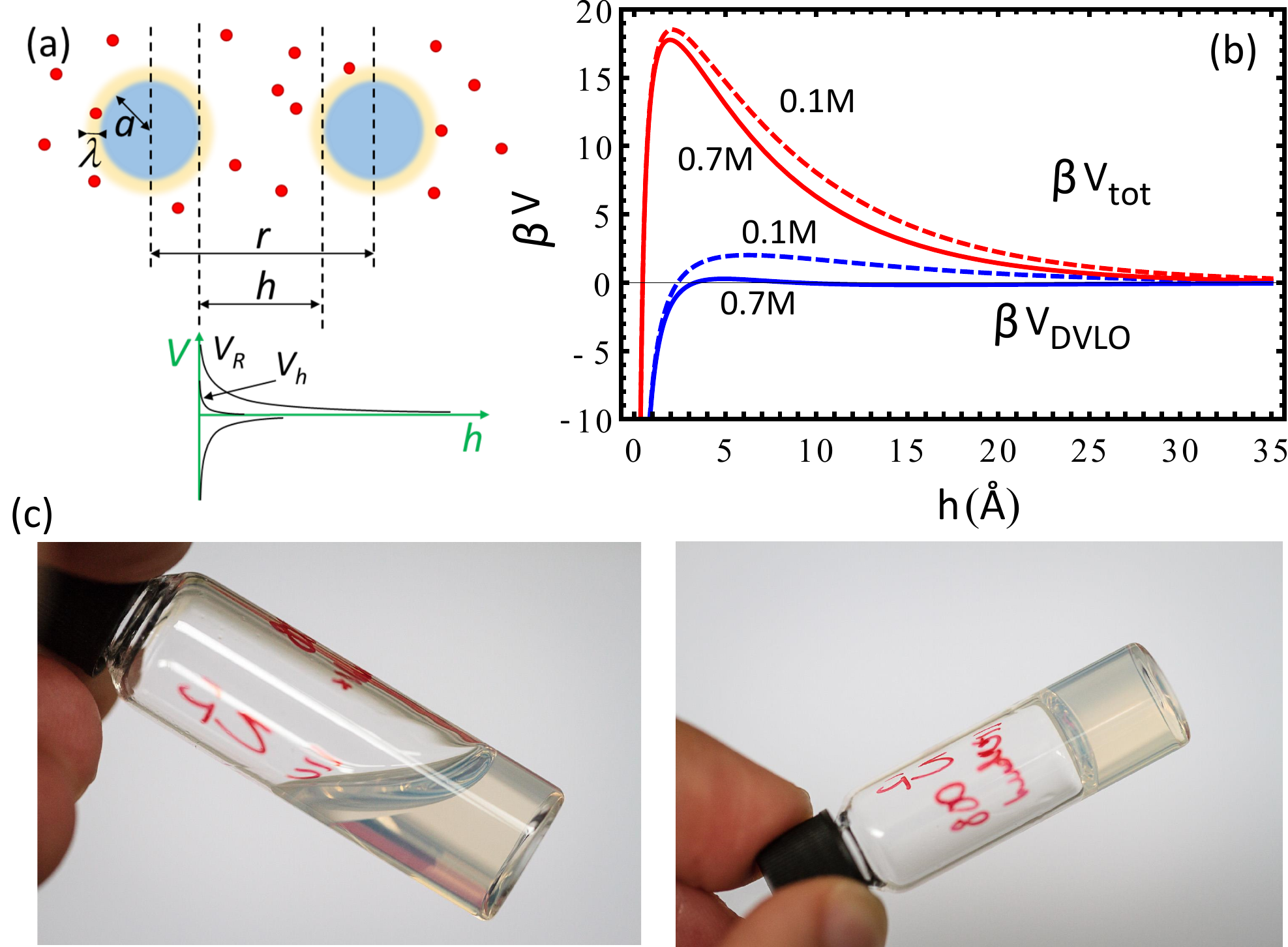}
\caption{ {The time required for gelation of the sample is related to the interparticle interaction potential, and the specific hydration of the ions has a huge effect, controlling the inter-colloid hydration repulsion at a  $h<\lambda$. (a) Schematic, and (b) plot, of the ``total'' interaction obtained in this work, including the strongly ion-specific short-ranged repulsive hydration potential. Also illustrated are the classical terms in the  DVLO interaction. At  very short distances, on the order of the hydrated ionic diameter, the repulsive shoulder dominates the interaction. (c) The macroscopic sample initially  flows as the
vial is inverted (left), while after some time  it becomes solid (right). This sample has $\phi_{\mathrm{Ludox}}=0.140$  and {374}\,mM \ce{NaCl}.}}
\label{fig:1}
\end{figure*}

 DLVO theory   predicts
 that there is a salt concentration at which the suspension aggregates (critical coagulation
 concentration); here the maximum DLVO interaction and its derivative
 are both zero (i.e. there is no energy barrier against aggregation).
 This concentration is proportional to $z^{-6}$, where $z$ is the valency of the salt
 ions, and
 is one of the successes of DLVO theory~\cite{Israelachvili2011a,Everett1988}. It is well known that DLVO theory breaks down completely for high salt concentrations (above 0.1M, which unfortunately is the regime of biological interest)~\cite{Bostrom2001,Parsons2011}. In this range, the basic assumptions of point charges, a solvent continuum and neglecting ion-surface adsorption and dispersion forces, are called into question. On a fundamental level, even the assumption that electrostatic forces and dispersion forces are additive is incorrect~\cite{Salis2014,Ninham1997}.


{It {is} shown in this letter that there is a spectacular failure of DLVO theory in
estimating gelation times, for identical particles, in the presence of different
monovalent salts \textit{even at low concentration}. {Changing the salt type dramatically affects the aggregation process}. Using a  model for relating the gelation kinetics to particle interactions, a non-DLVO hydration repulsion is characterised, and the dramatic changes in gelation kinetics are explained microscopically in terms of ion solvation and its interplay with the charged colloid surface. The proposed framework will make it possible to finely tune the gelation rate of nanoparticles simply by the choice of monovalent electrolyte species in the colloidal solution. }



 A minimum of  context and concepts proposed to explain the Hofmeister series are useful to the reader.    It  was first shown by Hofmeister~\cite{Kunz2004} that the stability of a colloidal solution (he made observations on proteins, which were then investigated by others~\cite{Piazza2000}, while other work investigated colloidal particles~\cite{Ninham1997,Huang2013,DosSantos2011,Paunov1996}) can be drastically different
 upon the addition of different salts of the same valency, even if all other parameters
 (such as the salt concentration) are kept constant.   Electrolytes could be arranged according to their efficiency in salting out
protein (now known as ``Hofmeister series''). The effect is understood to be related to how  salt ions  structure the water around themselves.
For monovalent cations the series
is \ce{NH_4^+}, \ce{Cs^+}, \ce{Rb^+}, \ce{K^+}, \ce{Na^+}, \ce{Li^+},
from most chaotropic (weakly hydrated, structure breakers) to most cosmotropic (strongly hydrated, structure makers),
and which extreme is
most destabilising depends on the surface properties
of the colloids~\cite{Lyklema2009}.

There are numerous, partially conflicting, theories as to the origin of these short-ranged ion-specific colloidal interactions, linking to the ioni size~\cite{Paunov1996,Paunov1999,Kralchevsky2011}.
%
%
%
Strongly polarisable ions are large and have more diffuse electron clouds. The energy penalty for being less well hydrated
(for example due to adsorption at an interface) is low for such ions because the charge can
be easily redistributed~\cite{Parsons2010}.
 The
   decreasing size trend  in going from chaotropic to cosmotropic in the Hofmeister series is consistent with this picture~\cite{DosSantos2011}.
    However the effective polarisability of the ions consists of contributions from both the ion itself and the solvent molecules in its hydration shell~\cite{Parsons2010}; This by extension may significantly augment the dispersion forces, and give further ion-specific interactions~\cite{Boroudjerdi2005}. Another suggestion is that the large electric field on the colloidal surface, arising from the finite size of the counterions, results in the ions acquiring appreciably large effective polarisabilities~\cite{Liu2014}.

Experiments with (negatively charged) mica surfaces showed that there is adsorption of cations.
More hydrated cations (such as \ce{Li^+}) are adsorbed only at high salt concentrations, while
the less hydrated ions adsorb at lower concentrations. However, once adsorbed, the cosmotropic
ions retain part of their hydration layer. This gives rise to a repulsive interaction as two
surfaces approach each other~\cite{Israelachvili2011a}. For mica, for example,  cosmotropic
ions are thus much more efficient at providing stabilisation
than chaotropic
ions.

Models have  been made to describe ion-specific distribution of ions near  surfaces, and their surface adsorption: an important factor
 is  the ion diameter, and whether or not the ions are hydrated~\cite{DosSantos2011}.
 Chaotropic ions have smaller effective diameters (as they are not hydrated), and they can adsorb to the surface.
 Based on this model
  it was possible to  calculate the critical coagulation concentration for a range of
  salts~\cite{DosSantos2011}.


\begin{figure}[t!]
\centering
\includegraphics[width=8cm]{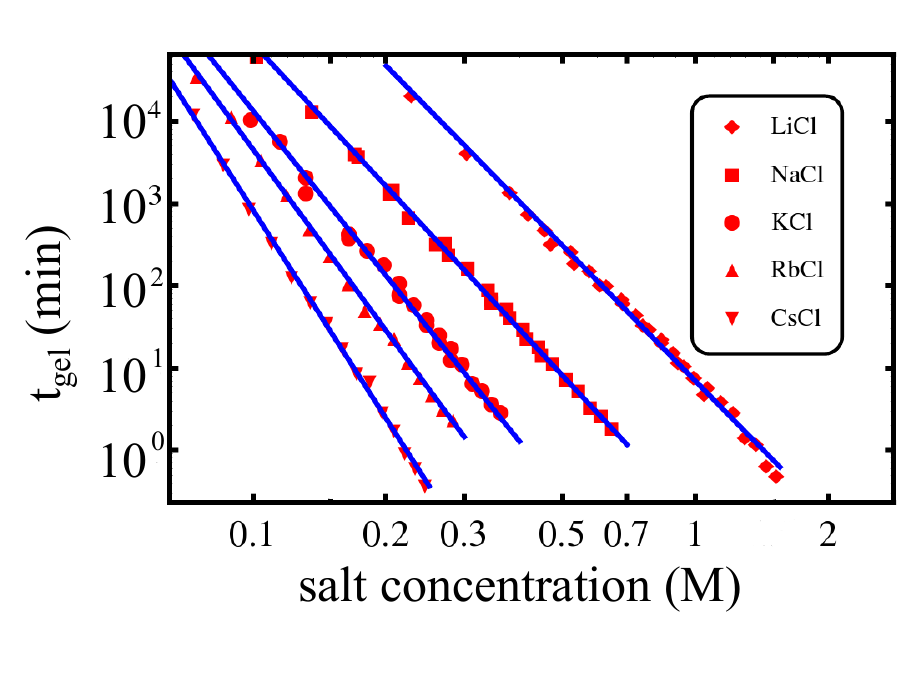}
\caption{ There is a very strong power law dependence (approximately -6 exponent) between gelation time and salt concentration, and a striking difference between the five monovalent salt species. The experimental observations (markers) are well recapitulated by  the theoretical predictions (one parameter fits,  as described in the text) (solid lines).  {Data is obtained from samples with volume fractions 0.13 and 0.14, and are undistinguishable. For each
curve, the only free fit parameter is the hydration force amplitude $F_{0}$, which is a function dependent on the salt-concentration   (see Fig.~\ref{fig1_br}). The theory is calculated assuming volume fraction 0.133, and provides a  match with the data for  values of $F_{0}$ well within the typical range of $10^{6}$ to $5 \times 10^{8}$ Nm$^{-2}$~(ref~\cite{Jia2006}).}}
\label{fig:3salts}
\end{figure}

 A fully quantitative description of hydration interactions  should include all  factors outlined above (finite ion sizes,  discrete nature of the solvent, many-body dispersion forces and polarization effects);
Such a description does not exist, and is beyond the scope of this letter, which instead aims to provide motivation, guidance and useful constraints for  future  models.
We take a simplified approach to modeling the hydration interactions, which is described below, and we show how the total interparticle interaction can be related to the gelation times, and thus measured.
The DVLO potential $V_{\mathrm{DVLO}}$ is a linear superposition of an attractive
van der Waals potential $V_{\mathrm{vdW}}$, an electrostatic repulsion potential $V_{\mathrm{R}}$
and a short-ranged Born repulsion potential $V_{\mathrm{B}}$, which we neglect in this study due
to the fact it does not impact on the gelation process. The attractive component is given by
\begin{eqnarray}
&&V_{\mathrm{vdW}}(r)=\frac{-A_{H}(r)}{6}\bigg(\frac{2a^{2}}{r^{2}-4a^{2}}+\frac{2a^{2}}{r^{2}}+\log\frac{r^{2}-4a^{2}}{r^{2}}\bigg),\nonumber\\
\end{eqnarray}
where $a$ is the colloid radius and $r=2a+h$ is the colloidal centre-to-centre separation,
see Fig.~\ref{fig:1}(a). The Hamaker function $A_{H}$
can be written  in the form~\cite{Vanni2002}
$A_{H}(r)=A_{\epsilon=0}f_{scr}(r)+A_{\epsilon>0}f_{ret}(r)$,
where $A_{\epsilon=0}$ is the zero frequency contribution which is screened by the counterions
through the screening function $f_{scr}(r)$, and  $A_{\epsilon>0}$ is the non-zero frequency
contribution which is mitigated by retardation through the retardation function $f_{ret}(r)$
(their full form is given in SI for completeness).

{The DLVO theory is based on a number of assumptions,  the most important of which is the linearized Poisson-Boltzmann treatment of the electric-double layer repulsion, which is valid only within the Debye-Hueckel limit for the surface potential, i.e. for potentials lower than 20-25\,mV. In our calculations, however, we used an extension due to Sader, Carnie and Chan~\cite{Sader1995}, which extends the validity of DLVO theory to much higher potentials:}
\begin{equation}
V_{\mathrm{R}}=4\pi\epsilon_{0}\epsilon_{m}\bigg(\frac{k_{B}T}{e}\bigg)^{2}Y(r)^{2}\frac{a^{2}}{r}\ln[1+\exp(-\kappa h)],
 \label{eq_Vdlvo}
\end{equation}
where $\epsilon_{0}$ is the permittivity of free space, $\epsilon_{m}$
is the relative permittivity of water, $k_{B}$ is the Boltzmann constant, $e$ is the counterion charge and $T$ is the temperature. The function $Y(r)$
is given in full in SI, and depends on the surface potential  $\psi_{0}$.


{The other assumptions of the theory  are the following: (a)~The ions are treated as point-like (their finite volume and excluded-volume effects are neglected); (b)~Spatial correlations among ions are neglected; (c)~Ion-adsorption on the colloid surface is neglected; (d)~Dissociation equilibria between charged species on the colloid surface and ions in solution are neglected. }
Many studies have shown that the interparticle potential deviates significantly from DLVO theory
below a surface-to-surface separation $h$ of about 2\,nm in
water~\cite{Israelachvili1996}. An additional repulsive potential
has been postulated to arise from the hydration of the water due to the presence of counterions
and/or on strongly hydrophilic surfaces. This potential is still not fully understood microscopically
and, as outlined above, there are different competing theories relating to its origins exist in the literature. {Since the 1970s there has been a general agreement that the {effective hydration} potential decreases exponentially from the surface~\cite{Marcelja1976}, and can thus be taken to have the following general form:}
\begin{equation}
V_{\mathrm{h}}=F_{0}\pi a\lambda^{2}\exp\bigg(-\frac{h}{\lambda}\bigg), \label{eq_Vh}
\end{equation}
where the fitting parameters $F_{0}$ and $\lambda$ control the magnitude and the decay of the potential, respectively.
This expression has been used quantitatively to successfully describe the huge energy barrier contributed by repulsion between structured water layers on hydrophilic surfactant-coated colloids, a big effect which cannot be explained by DLVO-theory alone~\cite{zaccone2008}.

The full potential we consider is then a sum of dispersion, electrostatic and surface hydration terms:
\begin{equation}
V_{\mathrm{tot}}\,=\,V_{\mathrm{vdW}}\,+\,V_\mathrm{R}\,+\,V_{\mathrm{h}}.
\label{eq:fullpot}
\end{equation}

The  interaction potential and gelation time are linked: A recent theoretical study~\cite{Zaccone2013} has established that the gelation time $t_{gel}$ can be
evaluated according to the following expression:
\begin{equation}
t_{gel}=\frac{1}{2k_{c}[(1+c)\phi_0/2]^{d_{f}/3}}, \label{eq_tgel}
\end{equation}
where $d_{f}$ is the fractal dimension of the clusters (which for reaction limited aggregation RLCA
is $d_{f}=2.1$~\cite{weitz90}), and $\phi_{0}=(4/3)\pi a^{3}n_{0}$ is the volume fraction of the colloids where
 $n_{0}$ is the total number of colloidal particles per unit volume, $n_{0}=N/V$. The parameter
  $c$ equates to $c=(1-\phi_{c})/\phi_{c}$, where $\phi_{c}$ is the critical volume fraction at
  which the systems gels (the zero-shear viscosity diverges), which for spherical-like clusters
  is $\phi_{c}\approx0.64$. The characteristic aggregation rate $k_{c}$ is given by the relation
\begin{equation}
k_{c}=\frac{n_{0}k_{agg}}{2},\label{eq:kc}
\end{equation}
where $k_{agg}$ is the rate constant of aggregation.
The stability ratio $W$ is defined as~\cite{Fuchs1934,Mewis2012}:
\begin{equation}
W=\frac{k_{S}}{k_{agg}}=2a \int \limits_{0}^{\infty} \,\frac{\exp(\beta V_{\mathrm{tot}})}{(2a+h)^{2}G(h)} \, \mathrm{d}h, \label{eq_W}
\end{equation}
where $V_{\mathrm{tot}}$ is the total interparticle potential, $k_{S}$ is the Smoluchowski diffusion
limited aggregation rate $k_{S}=(8/3)k_{B}T/\mu$ with $\mu$ the solvent viscosity. The
hydrodynamics of two spheres approaching is given by
$G(h)=({6(h/a)^{2} + 4(h/a)})/({6(h/a)^{2} + 13(h/a) + 2})$.
Combining Eqs.~\ref{eq:kc} and~\ref{eq_W}, and rewriting the colloidal concentration $n_{0}$ in terms
of the volume fraction $\phi_{0}$ and the volume of one particle $V_{p}$, we can recover the expression
\begin{equation}
k_{c}=\frac{4\phi_{0} k_{B}T}{3V_{p}W\mu}. \label{eq_kagg}
\end{equation}
This identity can be inserted into Eq.~\ref{eq_tgel} to obtain an explicit form for the gelation time
as a function of the sample material characteristics and interaction parameters:\\
\begin{equation}
t_{gel}=\frac{3V_{p}W\mu}{8\phi_{0} k_{B}T[(1+c)\phi_{0}/2]^{d_{f}/3}}.
\end{equation}

\begin{figure}[t!]
\includegraphics[width=8cm]{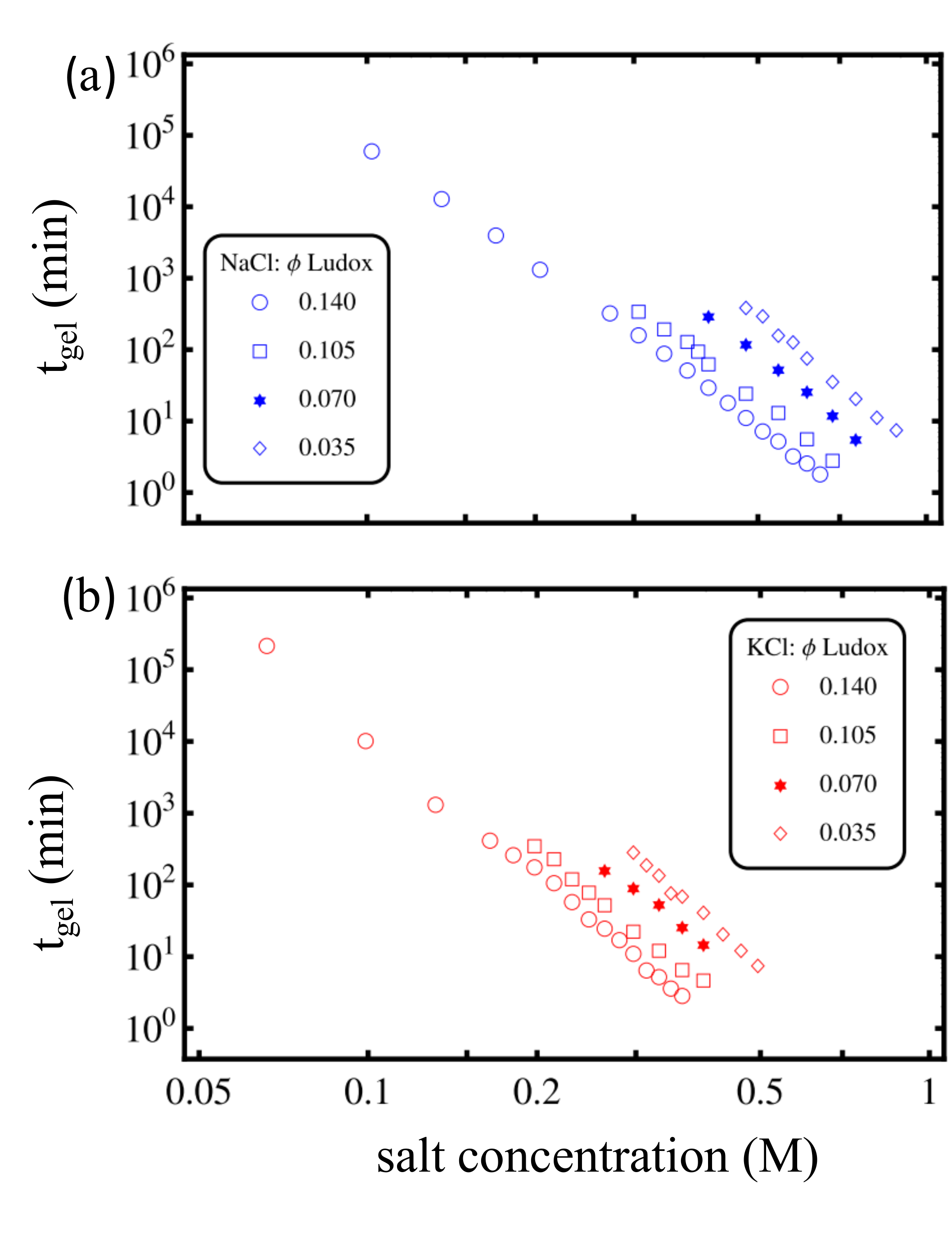}
\caption{The power law dependence of gelation time with salt concentration, and the strongly salt-specific  gelation times (shown here are (a)~\ce{NaCl} and (b)~\ce{KCl}), are seen in samples with varying Ludox concentration. Values of the best fit power law exponents are listed in Table~\ref{tab:exponents_saltconc}.\\
\label{fig:4Ludox}}
\end{figure}

Gelation times of various samples, varying salt type in~Fig.~\ref{fig:3salts}, salt concentration and Ludox concentration in Fig.~\ref{fig:4Ludox},  were
determined by checking when macroscopic samples no longer flowed.
Clearly, there is a progressive shortening of gelation times switching  samples with the same salt concentration, going  from \ce{LiCl} to \ce{NaCl},  \ce{KCl}, \ce{RbCl}  and \ce{CsCl},  and this is in agreement with the Hofmeister series. There are
\textit{four orders of magnitude} of difference in the gelation times - quite remarkable! -  for  the same concentrations of monovalent salts.

There is a clear power law dependence of the gelation times on the salt concentration, with exponents
(see Table~\ref{tab:exponents_saltconc}) all close to $-6$. There appears to be a
decrease of power law exponent (to a more negative value) as the cation becomes more chaotropic. For
samples with \ce{NaCl},  the power law exponent decreases (becomes more negative)
as the Ludox concentration decreases, but no clear trend is observed for samples with \ce{KCl}.
Reerink and Overbeek~\cite{Reerink1954} found similar power law behaviour for AgI colloids, with power law exponents
from around $-6$ to  $-11$, depending on the particle size and surface potential. Other power
law exponents had been reported in classical literature~\cite{Reerink1954},  ranging from $-2$ to
$-12$. The value of $-6$  falls into this range,
and seems very robust in our silica colloid data.\\

\begin{table}[t!]
\caption{Power law exponents for gelation times, from the  $t_{\text{gel}}$ vs. salt concentration data in
Figs.~\ref{fig:3salts} and~\ref{fig:4Ludox}; error quoted is the {95}\% confidence interval.       }
\label{tab:exponents_saltconc}
\begin{tabular}{ccc}
Salt & Volume fraction Ludox  & Exponent\\
KCl & $0.140$ & $-6.27 \pm 0.24$\\
KCl & $0.105$ & $-6.42 \pm 0.34$\\
KCl & $0.070$ & $-5.92 \pm 1.06$\\
KCl & $0.035$ & $-7.03 \pm 0.42$\\
NaCl & $0.140$ & $-5.67 \pm 0.11$\\
NaCl & $0.105$ & $-6.06 \pm 0.30$\\
NaCl & $0.070$ & $-6.47 \pm 0.56$\\
NaCl & $0.035$ & $-6.60 \pm 0.34$\\
LiCl & $0.133$ & $-5.49 \pm 0.07$\\
NaCl & $0.133$ & $-5.77 \pm 0.10$\\
KCl & $0.133$ & $-6.68 \pm 0.15$\\
RbCl & $0.133$ & $-7.33 \pm 0.06$\\
CsCl & $0.133$ & $-8.44 \pm 0.10$\\
\end{tabular}
\end{table}

\begin{figure}[t!]
\centering
\includegraphics[width=8cm]{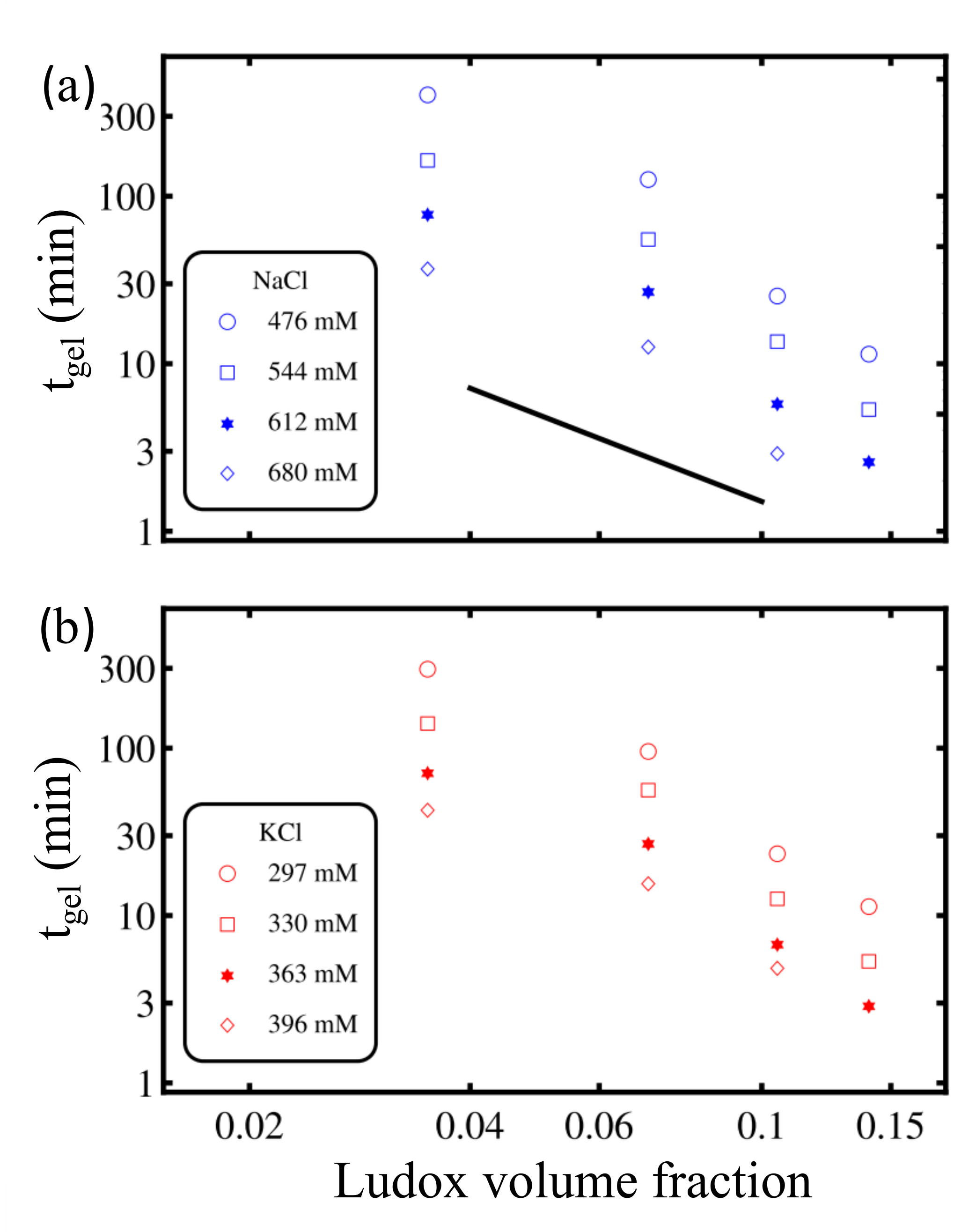}
\caption{Gelation times also depend on the  volume
fraction $\phi_{\mathrm{Ludox}}$ of Ludox particles: (a) \ce{KCl} and (b) \ce{NaCl}.
The solid line is a guide to the eye, illustrating the slope from a  power law with exponent -1.7, which is expected  for a fractal dimension of $2.1$.}
\label{fig:constSaltconc}
\end{figure}

 A power law relation is also expected if  the Ludox volume
 fraction is varied, at fixed  salt concentration~\cite{Zaccone2013}. If the colloidal aggregation is taking place with fractal
 dimension $d_f=2.1$ ({as is the case in the RLCA regime}), then the exponent in this plot is expected to be $-(d_f/3+1) = -1.7$;
 the data in
 Fig.~\ref{fig:constSaltconc} show a good agreement with this theoretical expectation. Note however that
 since (as explained later) we cannot assume to  have a constant hydration force,
 we  can't use the data of Fig.~\ref{fig:constSaltconc} to robustly go backwards and extract the
 fractal dimension.\\

The theoretical framework outlined  above, along with
the experimental data on the gelation times  at varying salt concentration,  now {allows us} to obtain the hydration parameters of the monovalent salts  LiCl, NaCl, KCl, RbCl and CsCl.
The first question to address is how to define the exponential decay length $\lambda$ in eq.~\ref{eq_Vh}. Throughout
the literature this parameter has been varied within the range $0.2-1.0$\,nm~\cite{Jia2006} for different colloidal systems.
 It seems reasonable to us to set $\lambda$ as the characteristic
hydration diameter of the counterions ( Cs$^{+}$ = 658\,pm, Rb$^{+}$ = 658\,pm,  K$^{+}$ = 662\,pm, Na$^{+}$ = 716\,pm,  Li$^{+}$ = 764\,pm)~\cite{Conway81}.\\

 The second, more delicate question, is how to determine the two other unknown interaction
 parameters, which are the amplitude $F_0$ of the hydration interaction (in eq.~\ref{eq_Vh}), and the surface potential $\psi_0$ (in eq.~\ref{eq_Vdlvo}).\\

\begin{figure} [t!]
\centering
\includegraphics[width=8cm]{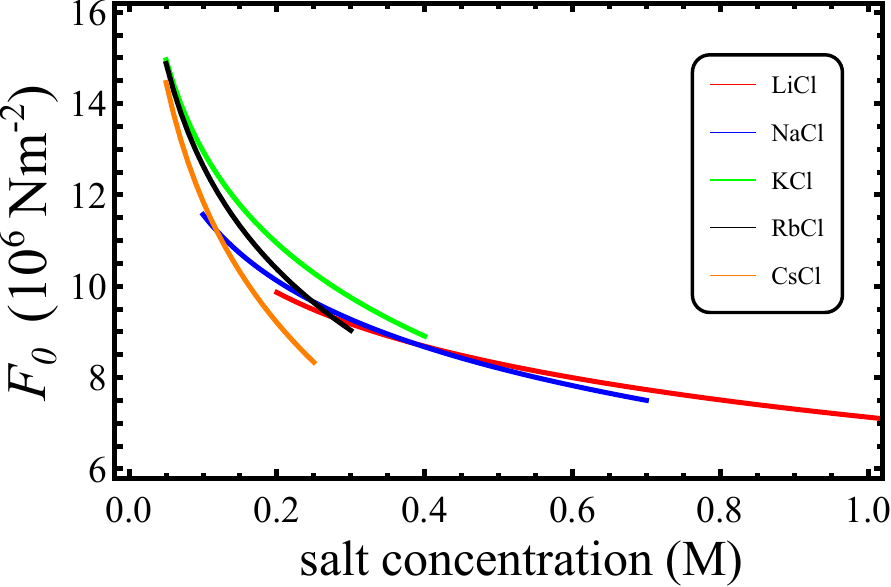}
\caption{The hydration force constant $F_{0}$ diminishes with increasing salt concentration. These values are obtained from  samples with a volume fraction $\phi_{\mathrm{Ludox}}$ of $0.133$.}
\label{fig1_br}
\end{figure}

\begin{figure}
\centering
\includegraphics[width=8cm]{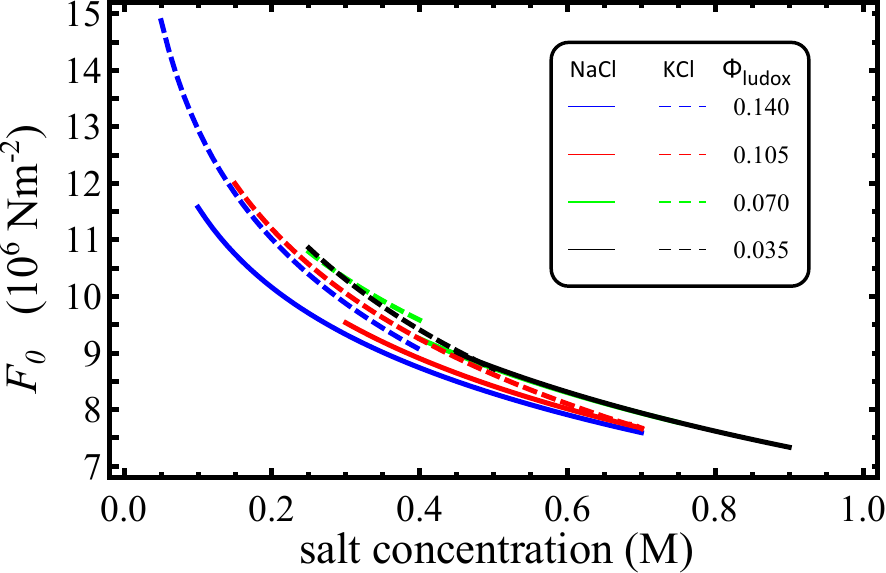}
\caption{The hydration force constant $F_{0}$ has a weak but systematic dependence on the particle volume fraction $\phi_{\mathrm{Ludox}}$.}
\label{fig3_br}
\end{figure}

Our first approach was to fix the hydration force constant $F_{0}$ {equal to a reasonable value from the literature}~\cite{Israelachvili2011a}, and allow the surface potential to vary as a function of the salt concentration. {Indeed, it might be expected that the association of counterions with silica surface groups will diminish the magnitude of the surface potential with increasing salt concentration, thus reducing the electrostatic repulsion and speeding up the gelation process~\cite{parsegian71}. This effect, while certainly present, is however far too small to justify by itself the rapid fall in the gelation times for increasing salt concentrations, as observed in Figs.~\ref{fig:3salts} and~\ref{fig:4Ludox}.} Coupled with the fact that experimental $\zeta$-potential measurements have been observed to be relatively insensitive to counterion adsorption on the surface~\cite{Merk2014}, we proceeded to approximate the surface potential to be constant, and set it equal to its dilute value of -30\,mV~\cite{Trompette2003}.
The second approach was therefore to proceed with the surface potential fixed, so that we could fit the experimental data in Figs.~\ref{fig:3salts} and~\ref{fig:4Ludox}, to obtain the hydration force constants $F_{0}$ {for each salt, and for each concentration}. This  is shown in Figs.~\ref{fig1_br} and~\ref{fig3_br}, as function of salt concentration and particle volume fraction. {It is clear from this framework that the reduction in $F_{0}$, and by extension in the repulsive hydration potential, {upon increasing the salt concentration, is what controls} the power-law relationship observed in Figs.~\ref{fig:3salts} and~\ref{fig:4Ludox}.}\\

{The  hydration {force magnitude values} $F_0$, are obtained at each salt concentration (c) from {a one-parameter fit} to the experimental data points in Fig.~\ref{fig:4Ludox}. The resulting functions $F_0(\phi(c))$  are shown in Fig.~\ref{fig1_br} (five salts) and Fig.~\ref{fig3_br}  (four different values of the particle
volume fraction $\phi_{\mathrm{Ludox}}$). Three trends are visible: (1)~$F_0(c)$ is higher for KCl than for any of the other ions, this means a non-monotonic behaviour in terms of ionic size;  (2)~$F_0(c)$ increases as Ludox concentration decreases; (3)~$F_0(c)$ decreases as salt concentration increases.   If our effective potential is valid, we have to assume all three effects are related to the association of the counterions with silica surface charge groups (or else, other factors might be contributing to the interaction, and are being assimilated into these $F_0$ trends).}

\begin{figure}[t!]
\centering
\includegraphics[width=8.2cm]{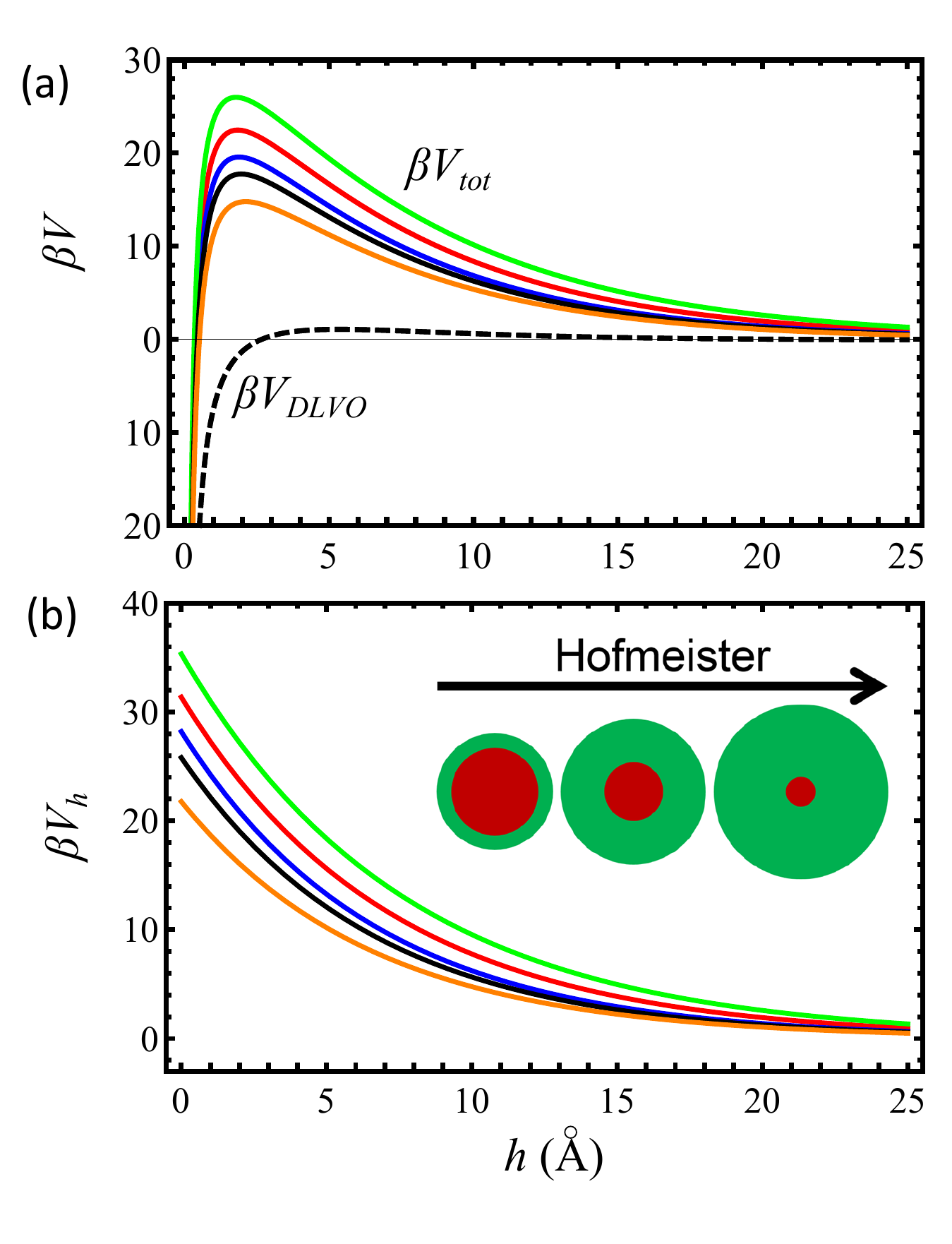}\\
\caption{The  salt specific repulsion is very important at short range, as shown in   (a)~by comparing  the total interparticle potential  ($V_{\mathrm{tot}}$, solid line) to  the potential  ($V_{\mathrm{DVLO}}$, dashed line)
without the hydration potential $V_{\mathrm{h}}$, and  in
  (b)~by plotting the hydration contribution by itself. The potentials decrease with salt species in the order LiCl-NaCl-KCl-RbCl-CsCl.   Salt
concentration is equal to $0.3$M for both plots.}
\label{fig4_br}
\end{figure}

 A plausible expected behaviour is that it is more difficult for the smaller, more hydrated monovalent cations to approach and thus associate with the surface  hydroxyl groups. Counterions adsorbed on the silica surface
 act as a repulsive force between particles. This explains why the hydration potential $V_{h}$ increases monotonically as a function of the hydration diameter (see Fig.~\ref{fig4_br}). Note that {since $V_{h} \propto F_{0}\lambda^{2}$ (eq.~\ref{eq_Vh}),  the trend of $F_{0}$   (Figs.~\ref{fig1_br} and~\ref{fig3_br}), which is observation (1) above, is more complex.} A greater {particle volume fraction corresponds to a greater total surface area, which is  consistent with diminishing counterion association per unit surface area, and  slight reduction of $F_{0}$, which is observation (2) above.} Also consistent with this picture is the fact that as we increase the salt concentration, the proportion of surface charge groups remaining free for counterion association dwindles and the ion-specific values of $F_{0}$ begin to converge. Fig.~\ref{fig4_br} shows that the hydration potential increases as the hydration diameter $\lambda$
lengthens, with the ordering of the salts by the relative strength of repulsive
hydration forces remaining the same over all distances. Therefore the greater the hydration
diameter of the counterion,  the more long-ranged
the hydration force becomes, and the greater the potential barrier to gelation. As a key result, the gelation time increases together with the hydration
diameter of the ions in the Hofmeister series.

\begin{figure}[t!]
\centering
\includegraphics[width=8cm]{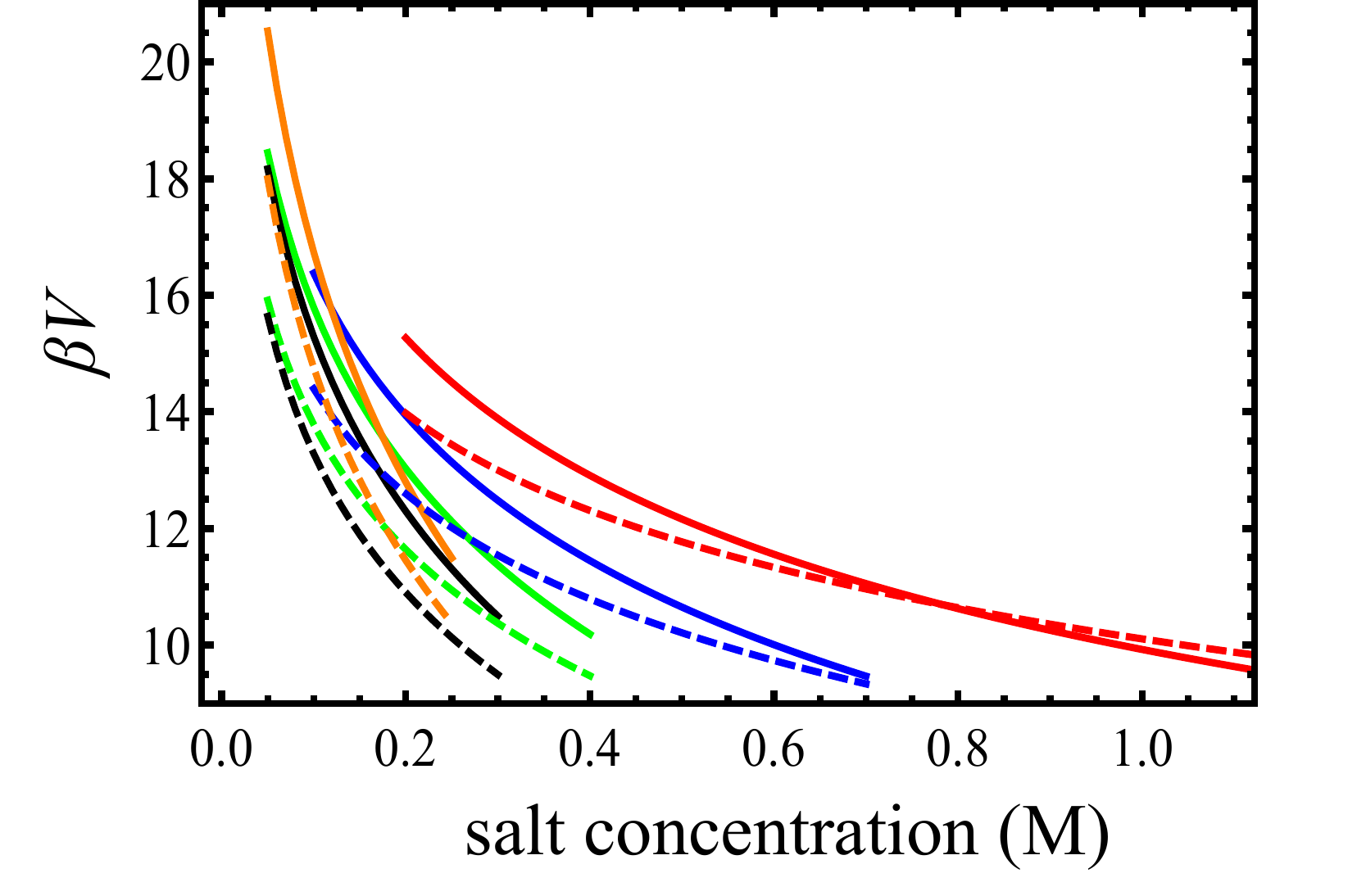}
\caption{With increasing salt concentration, the hydration potential decreases.  The interparticle separation is set here at the hydration diameter for each salt species. Solid lines are $\beta V_{tot}$ and dashed lines are $\beta V_{h}$, same color scheme as in~\ref{fig4_br}.}
\label{fig5_br}
\end{figure}

In Fig.~\ref{fig5_br}, the {hydration potential is observed to be a short-ranged} monotonically decreasing function of the salt concentration (observation (3) considered above). The precise origin of this effect is not obvious. In our simple picture of counterion adsorption we expect more ions to adsorb, the higher the bulk concentration.
 A possibility, similarly to what proposed in~\cite{Parsons2011}, is that one needs to consider a  loss of hydration shell when many ions adsorb, with a corresponding  decline in repulsion.
%
A final point we should remind the reader is that the surface charge density in the DLVO terms has been kept constant;  this is unlikely to be strictly correct, but it is very difficult to do otherwise with the data {at hand}~\cite{parsegian71}.
{Also, there are no experimental techniques to accurately evaluate the surface potential, and standard zeta-potential measurements are not adequate for this task. }
 So the question of whether this particular trend originates from some physical force or change in conditions at the gap, or an external force such as the bulk osmotic pressure, {remains to be properly addressed in future studies}.

Of particular interest to this work, Trompette and co-workers have described in a series of papers~\cite{Trompette2003,Trompette2004,Trompette2005}
 the effect of \ce{NH_4^+} and \ce{Na^+} on the stability of colloidal silica, and found that samples
with \ce{NH_4^+} aggregated much faster than those with \ce{Na^+}. They ascribed this to the different degrees of
hydration of the ions.
For the experiments described in this paper, only  \ce{LiCl},  \ce{NaCl}, \ce{KCl}, \ce{RbCl} and \ce{CsCl}  were studied since
 the ammonium salt ion is somewhat acidic, and the $p\mathrm{H}$ itself also affects stability of colloidal silica. \\

{Experimental papers investigating short-ranged hydration forces have classically employed Surface Force Apparatus (SFA), and more recently also Atomic Force Microscopy (AFM). These techniques can be applied to measure the forces between smooth solid surfaces, lipid bilayers and biomembrane surfaces~\cite{Israelachvili1996}. Some of the most recent studies include the use of AFM to estimate short-ranged hydration forces induced by multivalent salts~\cite{Ruiz-Cabello2014, Ruiz-Cabello2015}, the investigation of ion-specific Hofmeister effects between planar single-crystal sapphire~\cite{Lutzenkirchen2013} and the measurement of charge inversion as a function of pH and salt concentration.  These studies are in agreement~\cite{Franks2002,Jimenez2012} with the trends presented here,  whilst in other conditions the Hofmeister series is in reverse order~\cite{Morag2013}, but we have not found experiments that can be directly compared  to our results. We also note that our modeling shows that variations in the surface potential are insignificant compared to the dominant short-ranged hydration forces at close surface to surface separations.}
{Ion-specific double layer pressure has been calculated using the full nonlinear Poisson-Boltzmann equation with the addition of the ionic dispersion energy between the ions and the two interfaces~\cite{Bostrom2006}: By treating the electrodynamic ionic dispersion potentials on the same non-linear level as electrostatic potential, the  ion-specific Hofmeister effects are recovered. We believe that there is great scope to combine studies such as this with experimental approaches such as in this manuscript, to help isolate and calibrate the various possible mechanisms underpinning the hydration forces and bridge the gap between molecular and macroscopic observation.}

In conclusion, systematic experiments were carried out quantifying the gelation kinetics as a function of monovalent salt species. By using a simple model for the kinetics of cluster aggregation to fit experimental data of gelation times, at different salt conditions,  \textbf{for the first time} it was possible to extract the magnitude of a non-DVLO  hydration repulsion, that has a range set by the \textbf{solvated} diameter of the counterion.
 The very simple non-DVLO term used here is obviously coarse graining the detailed molecular  mechanisms ({the ordered ``rigid'' water layers repelling each other, and the energy required to ``squeeze'' these away as the particles approach contact}) and is an effective semi-empirical term. This approach is powerful because experiments can  then be readily fitted by a single  parameter, the amplitude of repulsion force ($F_0$, function of salt concentration). {The latter decreases with increasing salt concentration due to de-solvation of the counterions upon adsorption on the surface.
  The key finding  is  that the hydration repulsion correlates positively with the  {chaotropic nature} and the size of the cation species. This framework and the molecular-level mechanism proposed here can be used in the future to devise tunable gelation protocols of nanoparticles by choosing the salt type.}

\section{Methods}
\label{cpt:experimental}


All samples studied consist of colloidal silica, water and salt in different concentrations.
 Commercial silica colloids of Ludox HS-30, (Sigma-Aldrich), were used.
Ludox HS size measurements vary from 16.7\,nm~\cite{Chen2008}
to 18.5\,nm~\cite{Leysen2004} diameter, and we have taken 17\,nm diameter as a value in our calculations.

The Ludox was filtered prior to use, using Millipore Millex GS 0.22\,$\mu$m filters, to
 ensure the removal of larger aggregates. The resulting Ludox ``stock'' used in most of the experiments had a density
 of {1.23}\,g/ml and contained {31.5}\% silica by weight. The $p\mathrm{H}$ of the original
 Ludox suspension (before filtration and dilution) was $9.8$ (determined by the manufacturer), and
 was not regulated during the experiments. Stock solutions of salt were prepared
 using  \ce{LiCl}
 ({99.0}\%, Sigma Aldrich and 99\%, Acros),
 \ce{NaCl} ({99.0}\%, Sigma Aldrich and analysis grade, Merck), \ce{KCl} ({99.9}\%, Fischer and 99+\%, Acros),
 \ce{RbCl} (99.8+\%, Acros) and \ce{CsCl} (99+\%, Acros). All water used for preparation of stock solutions and samples
 was of Millipore grade.


The salt solution was always added last to make the final sample, of 1\,ml volume. All samples
were mixed on a vortex mixer for around 10~seconds immediately after the addition of salt stock.
 The gelation time was determined on a macroscopic scale, by gently inverting the vials to observe the presence
 (or not) of flow. When there was no discernable flow for around 1~second, the sample was judged to be gelled.
 This criterion is somewhat arbitrary, but was strictly adhered to, so that results of different samples are
 consistent. Fig.~\ref{fig:1}(c) shows
 photographs of a sample before and after its gelation time.

Preliminary runs were carried out, so that the gelation time was approximately known.
This ensured that subsequent samples were not inverted unnecessarily (we did not notice in any case correlations between the gelation time
and the frequency of inspection).
Identical samples prepared on different days did sometimes show different gelation times, which could be due
to small temperature changes or minor differences in sample composition or preparation. These differences were
never very large (around {10}\%) and do not affect the observed trends.



\textbf{Supporting Information Available}: Details the complete expressions for the attractive van der Waals potential and electrostatic
interaction terms used in this
work.
 This material is available free of charge via the Internet
http://pubs.acs.org.

\begin{acknowledgement}
We acknowledge useful discussions with {A.Routh and L.Cipelletti,   preliminary work by H.Hong, and financial  support from:   Unilever Plc (E.S.);     the
Ernest Oppenheimer Fellowship at Cambridge (to 1st
June 2014),  and by the Technische Universit\"{a}t M\"{u}nchen
Institute  for  Advanced  Study,  funded  by  the  German
Excellence Initiative and the European Union Seventh
Framework Programme under grant agreement 291763 (A.Z.);   the Winton Programme for the Physics of Sustainability (B.O.C.).} All the data needed to support and reproduce these results are presented within the paper.
\end{acknowledgement}



\providecommand{\latin}[1]{#1}
\providecommand*\mcitethebibliography{\thebibliography}
\csname @ifundefined\endcsname{endmcitethebibliography}
  {\let\endmcitethebibliography\endthebibliography}{}


\end{document}


\section*{The complete expression for the attractive van der Waals potential.}
This SI details, for completeness, the terms in eq.~1 of the main text~\cite{Vanni2002}:
\begin{eqnarray}
A_{\epsilon=0}&=&\frac{3}{4}k_{B}T\bigg(\frac{\epsilon_{p}-\epsilon_{m}}{\epsilon_{p}+\epsilon_{m}}\bigg)^{2} \nonumber\\
A_{\epsilon>0}&=&\frac{3h\sqrt{\nu_{p}\nu_{m}}}{64\bar{n}^{7/2}}*\nonumber\\
&&\,\,\frac{X^{2}\bar{n}^{2}+2X(n_{p}^{2}-n_{m}^{2})\bar{n}+(3+2Y)(n_{p}^{2}-n_{m}^{2})^{2}}{[(Y-\sqrt{Y^{2}-1})^{1/2}+(Y+\sqrt{Y^{2}-1})^{1/2}]^{3}}\nonumber\\
f_{scr}(r)&=&(1+2\kappa h)\exp(-2\kappa h)\nonumber\\
f_{ret}(r)&=&\bigg[1+\bigg(\frac{\pi Qh}{4\sqrt{2}}\bigg)^{3/2}\bigg]^{-2/3},
\end{eqnarray}
where $\kappa$ is the inverse Debye length,
\begin{equation}
\label{eq:DebyeLength}
\kappa^{-1} = \left( \frac{\varepsilon \varepsilon_0 k_{\text{B}}T}{e^2 \sum_i c_i z_i^2} \right)^{1/2}.
\end{equation}

{Here c denotes the salt concentration} and
\begin{eqnarray}
& \bar{n}=\sqrt{\frac{n_{p}^{2}+n_{m}^{2}}{2}}\nonumber \\
& X=\frac{\nu_{p}}{\nu_{m}}(n_{p}^{2}-1)-\frac{\nu_{m}}{\nu_{p}}(n_{m}^{2}-1)\nonumber \\
& Y=\frac{1}{4\bar{n}}\bigg[\frac{\nu_{p}}{\nu_{m}}(n_{p}^{2}+1)+\frac{\nu_{m}}{\nu_{p}}(n_{p}^{2}+1)\bigg]\nonumber \\
& Q=2\pi n_{m}(n_{p}^{2}-n_{m}^{2})^{1/2}\frac{\sqrt{\nu_{p}\nu_{m}}}{c}.
\end{eqnarray}
The subscript $p$ refers to the particle (silica) and the subscript $m$ refers to the solvent (water).
$n$, $\epsilon$ and $\nu$ refer to the refractive index, the dielectric constant and the main absorption
 frequency of the silica and the water.

\section*{The complete expression for electrostatic components.}

The electrostatic repulsion, beyond the limit of validity of the Debye-Hueckel approximation,
is given by the Sader-Carnie-Chan expression~\cite{Sader1995} as:
\begin{eqnarray}
&&V_{\mathrm{R}}=4\pi\epsilon_{0}\epsilon_{m}\bigg(\frac{k_{B}T}{e}\bigg)^{2}Y(r)^{2}\frac{a^{2}}{r}\ln[1+\exp(-\kappa h)]\nonumber\\
&&Y(r)=4\exp\left(\frac{\kappa h}{2}\right)\mathrm{arctanh}\left(\exp\left[\frac{-\kappa h}{2}\right]\tanh\left[\frac{\psi_{0} e}{4k_{B}T}\right]\right),\nonumber\\
 \label{eq_Vdlvo}
\end{eqnarray}
where $\psi_{0}$ is the surface potential, $\epsilon_{0}$ is the permittivity of free space, $\epsilon_{m}$
is the relative permittivity of water, $k_{B}$ is the Boltzmann constant, $e$ is the counterion charge and $T$ is the temperature.